# Turing Minimalism and the Emergence of Complexity

Hector Zenil
Department of Computer Science
University of Oxford, United Kingdom

Abstract:
**Not only did Turing help found one of the most exciting areas of modern science (computer science), but it may be that his contribution to our understanding of our physical reality is greater than we had hitherto supposed. Here I explore the path that Alan Turing would have certainly liked to follow, that of complexity science, which was launched in the wake of his seminal work on computability and structure formation. In particular, I will explain how the theory of algorithmic probability based on Turing's universal machine can also explain how structure emerges at the most basic level, hence reconnecting two of Turing's most cherished topics: computation and pattern formation**

Alan Turing established such a direct connection between the concept of the algorithm and a purely mechanical process that he left few doubts concerning the physical implementation and generality of his programmable machines. At the beginning of the twentieth century and through the end of the Second World War, computers were human, not electronic, mainly women. The work of a *computer* consisted precisely in solving tedious arithmetical operations with paper and pencil. This was looked upon as work of an inferior order.

At an international mathematics conference in 1928, David Hilbert and Wilhelm Ackermann suggested the possibility that a mechanical process could be devised that was capable of proving all mathematical assertions. This notion is referred to as *Entscheidungsproblem* (in German), or 'the decision problem'. If a *human computer* did no more than execute a mechanical process, it was not difficult to imagine that arithmetic would be amenable to a similar sort of mechanization. The origin of *Entscheidungsproblem* dates back to Gottfried Leibniz, who having (around 1672) succeeded in building a machine based on the ideas of Blaise Pascal that was capable of performing arithmetical operations (named *Staffelwalze* or the *Step Reckoner*), imagined a machine of the same kind that would be capable of manipulating symbols to determine the truth value of mathematical principles. To this end Leibniz devoted himself to conceiving a formal universal language, which he designated *characteristica universalis*, a language that would encompass, among other things, binary language and the definition of binary arithmetic.

In 1931, Kurt Gödel arrived at the conclusion that Hilbert's intention (also referred to as 'Hilbert's programme') of proving all theorems by mechanizing mathematics was not possible under certain reasonable assumptions. Gödel advanced a formula that codified an arithmetical truth in arithmetical terms and that could not be proved without arriving at a contradiction. Even worse, it implied that there was no set of axioms that contained arithmetic free of true formulae that could not be proved.

In 1944, Emil Post, another key figure in the development of the concepts of computation and computability (focusing especially on the limits of computation) found that this problem was intimately related to one of the twenty-three problems (the tenth) that Hilbert, speaking at the Sorbonne in Paris, had declared the most important challenges for twentieth-century mathematics.

Usually, Hilbert's programme is considered a failure, though in fact it is anything but. Even though it is true that Gödel debunked the notion that what was true could be proved, presenting a negative solution to the 'problem of decision', and Martin Davis (independently of Julia Robinson) used Gödel's negative result to provide a negative solution to Hilbert's tenth problem (the argument for which was completed by Yuri Matiyasevich), Hilbert's supposedly failed programme originated what we now know as Computer Science, the field that wouldn't have been possible without Alan M. Turing's concept of the universal machine.

## One machine for everything

Not long after Gödel, Alan M. Turing made his appearance. Turing contemplated the problem of decision in much cruder terms. If the act of performing arithmetical operations is mechanical, why not substitute a *mechanical device* for the *human computer*? Turing's work represented the first abstract description of the digital general-purpose computer as we know it today. Turing defined what in his article he termed an '*a*' computer (for 'automatic'), now known as a Turing machine.

A Turing machine is an abstract device which reads or writes symbols on a tape one at a time and can change its operation according to what it reads, and move forwards or backwards through the tape. The machine stops when it reaches a certain configuration (a combination of what it reads and its internal state). It is said that a Turing machine produces an output if the Turing machine halts, while the locations on the tape the machine has visited represent the output produced.

The most remarkable idea advanced by Turing is his demonstration that there is an '*a*' machine that is able to read other '*a*' machines and behave as they would for an input *s*. In other words, Turing proved that it was not necessary to build a new machine for each different task; a single machine that could be reprogrammed sufficed. This erases the distinction between program and data, as well as between *software* and *hardware*, as one can always codify data as a program to be executed by another Turing machine and vice versa, just as one can always build a *universal* machine to execute any program and vice versa.

Turing also proved that there are Turing machines that never halt, and if a Turing machine is to be universal and hence able to simulate any other Turing machine or computer program, it is actually expected that it will never halt for an infinite number of inputs of a certain type (while halting for an infinite number of inputs of some other type). And this is what Turing would have expected, given Gödel's results and what he wanted to demonstrate: that Hilbert's mechanisation of mathematics was impossible. This result is known as the *undecidability of the halting problem*.

In his seminal article Turing defined not only the basis of what we today know as digital general-purpose computers, but also software, programming and subroutines. And thus without a doubt it represents the best answer to date that we have to the question 'What is an algorithm?' In fact in Alan Turing's work on his universal machine, he even introduced the concept of a subroutine that helped him in his machine construction. These notions are today the cornerstone of the field that Turing, more than anyone else, helped establish, viz. Computer Science.

Once we approach the problem of defining what an algorithm is and arrive at the concept of universality that Turing advanced, the question to be considered in greater detail concerns the nature of algorithms. Given that one now has a working definition of the algorithm, one can begin to think about classifying problems, algorithms and computer programs by, for example, the time they take or the storage memory they may require to be executed. One may assume that the time required for an algorithm to run would depend on the type of machine, given that running a computer program on a *Pentium PC* is very different from executing it on a state-of-the-art super computer. This is why the concept of the Turing machine was so important-- because any answers to questions about problem and algorithm resources will only make sense if the computing device is always the same. And that device is none other than the universal Turing machine. So for example, every step that a Turing machine performs while reading its tape is counted as a time step.

Many algorithms can arrive at the same conclusion taking different paths but some may be faster than others, but this is now a carefully considered matter when fixing the framework for Turing's model of computation: one asks whether there is an algorithm that surpasses all others in economy as regards the resources required when using exactly the same computing device. These are the questions that opened up an entire new field in the wake of Turing's work, the development of which Turing would certainly have been delighted to witness. This field is today referred to as the theory of Computational Complexity, which would not have been possible without a concept such as that of the universal Turing machine. The theory of Computational Complexity focuses on classifying problems and algorithms according to the time they take to compute when larger inputs are considered, and on how size of input and execution time are related to each other. This is all connected to two basic resources needed in any computation: space (storage) and time. For example, one obvious observation relating to this theory is that no algorithm will need more space than time to perform a computation. One can then quickly proceed to ask more difficult but more interesting questions, such as whether a machine can execute a program faster if it is allowed to behave probabilistically instead of fully deterministically. What happens if one adds more than one tape to a universal Turing machine operation? Would that amount to implementing an algorithm to solve a problem much faster? Or one may even ask whether there is always an efficient algorithm for every inefficient algorithm, a question that may lead us to a fascinating topic connecting computers and physics.

## The world of simple programs

If algorithms can come in a variety of types, some slow, others faster, what is it that allows nature to produce, with no apparent effort, what seem to us to be complex objects? (see Figs. 4, 5 and 6). These range from the laws of physics to the formation of matter and galaxies to the beginning of life on Earth (and possibly in other parts of the universe). In the end, one can see all these natural phenomena as a kind of computation, regardless of whether it is of exactly the same type as that performed by a Turing machine. This latter possibility cannot be completely disregarded. Thanks to Turing we know that even simple devices such as universal Turing machines possess incredible power.

One of the natural world's most fascinating characteristics is that it presents a wide range of physical and biological systems that behave in different ways, just like algorithms, most of them having some regular features while nonetheless being hard to predict. Climate is a case in point. Even though it is cyclical, it is impossible to predict its details more than a week in advance.

Where does nature's complexity come from? Throughout human history we have encountered objects, in particular mathematical ones that seem complex to us. One set of such objects comprises numbers that can de expressed as the division $p/q$, with $p$ and $q$ being integers. Numbers 5, 0.5 or even infinite numbers such as 0.333… can be written as 5/1, 1/2, and 1/3, respectively. But as far back as the ancient Greeks numbers have been known, such as $\pi$ and the square root of 2, which cannot be expressed in this way. One could think of arithmetical division as an algorithm that can be performed by a Turing machine, the result being provided in the output tape. Multiplication, for example, is an algorithm to shorten the number of steps needed to perform the same operation using only addition. In the case of numbers that admit a rational representation $p/q$, the algorithm of the division of integers consists of the common procedure of finding quotients and remainders. In the case of numbers such as $\pi$ and the square root of 2, the algorithm produces an infinite non-periodic expansion, so that the only way to represent them is symbolically (i.e. $\pi$ and $\sqrt{2}$). The Pythagoreans found that those numbers with ostensible infinite complexity could be produced from very simple operations, for example, when seeking the value of the hypotenuse of a right triangle with sides of length 1. Since Euclid, it has also been known that such numbers are not the exception among real numbers that are found, for example, in the continuous interval (0, 1).

In algorithmic terms, rational and irrational numbers are different in nature. When one starts a Turing machine that implements the algorithm for division, there is no algorithm that allows for the production of an irrational number followed by halting, whereas the division of rational numbers can halt (when the remainder is zero) or enter an infinite cycle that will produce a repetitive decimal expansion.

In engineering, including systems programming, the intuition of what is complex (in comparison to an irrational number in mathematics, for example) has been radically different. The usual assumption has been that to produce something that seems complex, a process that is just as complex has to be devised. This issue, however, is closely connected to Turing's concept of universality, given that a universal Turing machine that is programmable is, in principle, capable of producing any degree of 'complexity', for example, the type of complexity (or randomness) that one can see in the decimal expansion of π.

If Euclid's algorithm for division or π can produce such apparent complexity, how usual is it to run a random computer program that produces the same complexity? If computer programs that produce complexity need a very complex description, the probability of finding one small enough would be very low. For example, even though Turing's 1936 article contains all the main elements of the traditional description of a universal Turing machine capable of reproducing the type of complexity to be found in the digits of π, the construction of his universal machine requires at least 18 states, and at least 23 instructions (the exact number cannot be calculated on the basis of Turing's article due to the fact that he uses subroutines that can be implemented on machines of different sizes).

Whatever the actual threshold for reaching Turing universality, it had typically been thought to be high (a case in point: von Neumann's universal builder, a system that was the anticipation of the modern concept of cellular automata, requires 29 states), and it was thought that a universal machine would require a certain minimum *complexity* (at least as to the number of states and symbols required to describe it). In an experiment with extremely small and simple computer programs, Stephen Wolfram found that this threshold of complexity and universality was likely to be extremely low, and that very little was required to find a machine that produced high complexity or that was capable of being Turing universal. Wolfram's computer programs called *Elementary Cellular Automata* with rule numbers 30 (Fig. 1) and 110 (Fig. 2) are the best examples of rich behaviour obtained from minimalistic computer programs.

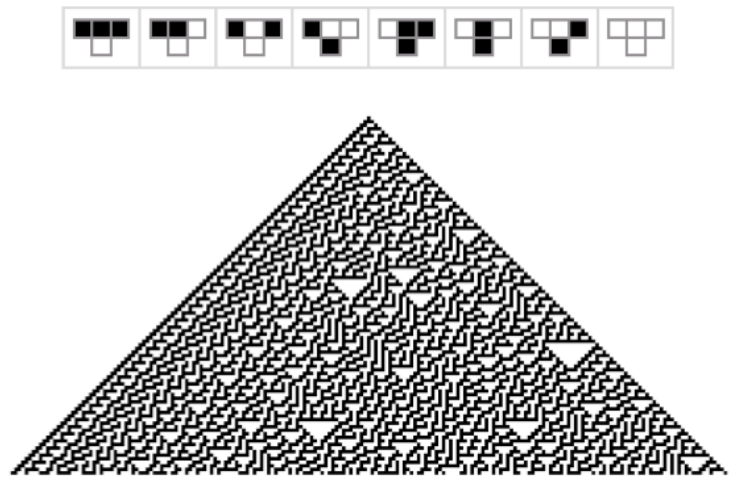

Figure 1: Evolution of a simple program called Rule 30 (100 steps are shown in this image) that, surprisingly, generates apparent complexity in spite of starting with a single black cell (the simplest initial condition for this computer program). The icon at the top of the program's evolution shows the transition table. Rule 30 is, by perhaps any standard, the simplest computer program that produces this degree of apparent disorder.

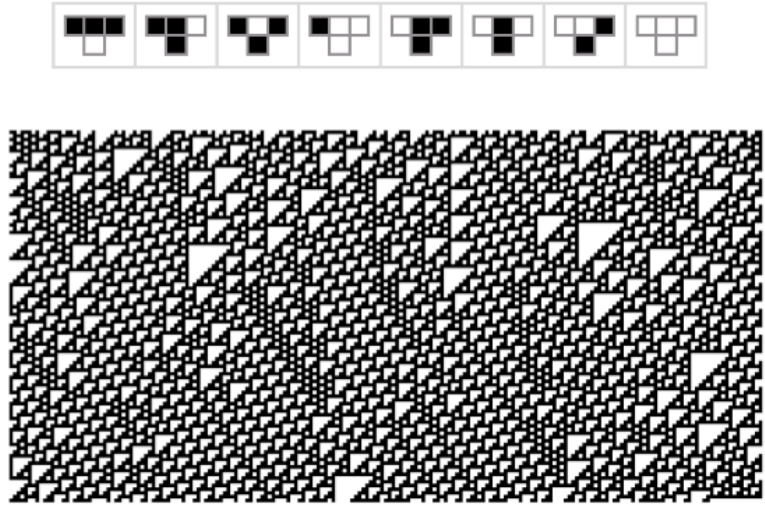

Figure 2. Evolution of a simple program called Rule 110 (100 steps shown) is capable of Turing universality due to the presence of persistent structures--some of which can be observed in this figure--that allow for the transfer of information. This means that a rule as simple as this (the rule appears at the top) can compute any function that a Turing machine can compute.

Certain parts of the universe seem ordered and structured. On Earth, for example, life is an example of organization and structure, contrasting with what can be deemed background noise (comparable to what appears on the screen of an old non-tuned analog television) left over by the Big Bang and serving as proof of the state in which the universe found itself after its first moments of existence. What Turing would probably never have guessed is that when running random Turing machines, the machines produce highly structured objects. Could this be merely an interesting analogy or could it perhaps be an actual indication that the universe is more algorithmic than initially expected? If so, then Turing machines and computer programs are not just products of the human imagination, they are perhaps responsible for the order in the universe. Alan Turing may have had an intuitive answer to this question, as he was also interested in structure formation and helped found another area in biochemistry called Morphogenesis. Turing was interested in pattern formation, starting from a simple shape which would first break its symmetries at random.

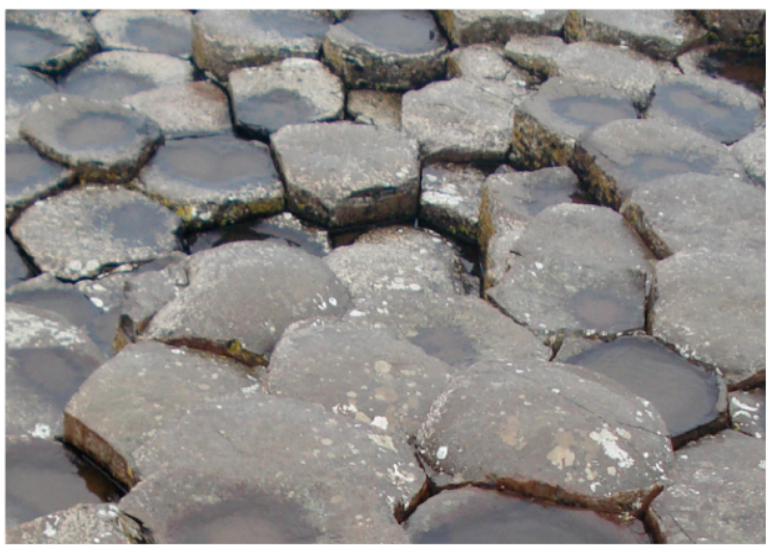

Figure 3. The Giant's Causeway in Northern Ireland. Hexagonal basalt columns formed by sea water cooling incandescent lava relatively quickly, around 60 million years ago, as a result of the interplay of a small number of forces. (Picture, H. Zenil, 2009)

Looking at Figs. 3 and 4, we may form the impression that the structures shown cannot possibly be the product of natural processes unassisted by human intervention. However, once we realise that, despite appearances, they are indeed structures that have been formed naturally, we accept that the process through which they have come into existence is neither exceptional nor complex, but simple and relatively common. In Fig. 3, hexagonal columns can be seen that were formed from accumulated lava that was rapidly cooled as soon as it came in contact with sea water. In Fig. 4 we see a great white circle on the ground, a result of the growth of mushrooms. How do the mushrooms communicate with each other to arrange themselves in a circle? The first mushrooms in a 'fairy ring' emerge in the middle, and start reproducing outwards. The older mushrooms in the center die off whereas the younger mushrooms form even larger circles. Traditionally it has been thought that objects with structure can only be produced by an intelligent mind. How can nature create this structure?

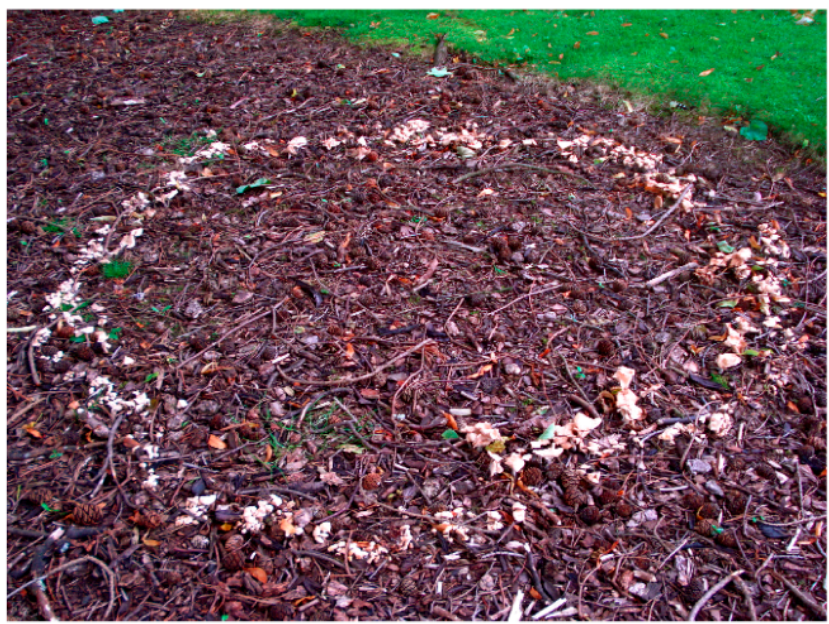

Figure 4. A 'fairy ring' on the ground is formed in a natural manner by the differential growth of Micelios mushrooms, i.e. an extremely simple mechanical process. (Picture, H. Zenil, Dundee, Scotland. 2012)

## An algorithmic theory of emergence

Just as the formulas for the production of the digits of π are compressed versions of π, the laws of physics can be seen as systems that *compress* natural phenomena. These laws are valuable because it is thanks to them that the result of a natural phenomenon can be predicted without having to wait for it to unfold in real time, e.g. one can solve the equations that describe planetary movement instead of waiting two years to know the future positions of a planet. For all practical purposes the laws of physics are like computer programs and the scientific models we have of them, executable on digital computers and susceptible of numerical solutions.

There is a measure that describes the probability of a universal Turing machine producing a string *s* when running a computer program produced at random. In the form of an equation this measure can be written as follows,

$$m(s) = \sum_{\{p:U(p)=s\}} 1/2^{|p|} \qquad (1)$$

In which |p| is the length (in bits) of the programs that produce the string *s* running on a universal Turing machine *U*. In order to work *U* has to fulfill a minimal technical requirement, viz. that no valid computer program is the beginning of another valid computer program. The measure *m* induces a distribution over the set of strings that is known as the *Universal Distribution*, the properties of which have even been described as 'miraculous' in the literature on computer theory.

The notion behind *m* is very intuitive. If one wished to produce the digits of π randomly, one would have to try time after time until one managed to hit upon the first numbers corresponding to an initial segment of the decimal expansion of π. The probability of success is extremely small: 1/10 digits multiplied by the desired quantity of digits. For example, $(1/10)^{2400}$ for a segment of 2400 digits of π. But if instead of shooting out random numbers one were to shoot out computer programs to be run on a digital computer, the result would be very different. A program that produces the digits of π would have a higher probability of being produced by a computer program. Concise and known formulas for π could be implemented as short computer programs that would generate any arbitrary number of digits of π.

It should be noted that the largest term in the sum of equation 1 is obtained when the denominator is the smallest, that is, when |p| is the smallest, namely the shortest length of program *p* in bits that produces *s*.

Not coincidentally, the length of the shortest program that produces a string is acknowledged as a measure of randomness called algorithmic or Kolmogorov complexity. The idea is relatively simple. If a string of length *s* cannot be produced by a program that produces *s* so that |p| < |s| where |p| denotes the length of the program *p* in number of bits, then the string *s* is considered random because it cannot be described in a shorter way than by *s* itself, there being no program *p* that generates *s* whose length is shorter than *s*. Kolmogorov complexity is defined as follows:

$$C_U(s) = \min\{|p|, U(p) = s\} \qquad (2)$$

The measure of algorithmic probability *m* assigns a low algorithmic complexity to the strings that are produced more often, whereas those with a lower frequency have a greater Kolmogorov complexity, that is to say, they seem more random.

The invariance theorem guarantees that the value of *C*, whether calculated with one particular universal Turing machine or any other universal Turing machine, is the same at the limit. Formally, if $U_1$ and $U_2$ are two universal Turing machines and $C_{U_1}(s)$ and $C_{U_2}(s)$ are the values for the algorithmic complexity of *s* for $U_1$ and $U_2$ respectively, there exists a constant *c* such that

$$|C_{U_1}(s) - C_{U_2}(s)| < c \qquad (3)$$

Thus, the longer the string, the less important *c* is and the more stable the Kolmogorov complexity value *C* is.

One of the disadvantages of *C* is that, given the halting problem for Turing machines, C is not computable, which is to say that given a string, there is no algorithm that returns the length of the shortest computer program that produces it. We have already seen how *m(s)* relates to *C(s)*, given that according to its definition from Eq. 1, *m* obtains the greater part of its value from the shortest program that produces *s*. The Coding theorem formalises this relationship:

$$C(s) \sim -\log m(s) \qquad (4)$$

The theorem indicates that the algorithmic complexity of a string *s* is very close (up to an additive constant) to the negative value of the logarithm of the frequency of s.

# Concluding remarks

Running random programs can be seen as the simulation of a hypothetical binary universe that, besides constituting an alternative to compression algorithms, affords us a framework for studying and comparing with the real world, with the way in which patterns in the universe are distributed. In a world of computable processes in which the laws (like programs) do not have a slantwise distribution, $m(s)$ would indicate the probability of a natural phenomenon occurring as a result of running a program. Distribution $m(s)$ has an interesting particularity: it can start from basically anything, and, like a robust distribution, it remains qualitatively unchanged. It is the process that determines the form of the distribution and not the initial distribution of the programs. This is important because one does not make any strong initial assumption about the distribution of the initial conditions or of the laws of physics.

Computer programs can be looked at from a certain vantage point as laws of physics. If one starts with a random initial condition (input) and executes a program chosen at random, there is a very good probability that its final appearance will be regular, and frequently very well organised. By the same token if one were to shoot out particles at random, the probability of groups forming in the way they do would be so small--in the absence of laws of physics--that nothing whatsoever would happen in the universe. Perhaps Alan Turing hasn't only helped us understand the world of computing machines and computing programs, founding an entire scientific area, but has also taught us a great deal about the universe in which we live and how it came to be the way it is.

# Further Reading